# FloatSOM: GPU-Accelerated, Distributed, Topology-Flexible Self-Organizing Maps

**Tony Xu** *tony.xu@anu.edu.au*
*John Curtin School of Medical Research, Australian National University*

**Sarah Klamt**
*Department of Information Technology and Electrical Engineering, ETH Zürich*

**Katherine Turner**
*Mathematical Data Science Centre, Australian National University*

**Anne Brüstle**
*John Curtin School of Medical Research, Australian National University*

**Felix Marsh-Wakefield**[†]
*Centenary Institute of Cancer Medicine and Cell Biology*
*University of Sydney*

**Givanna Putri**[†]
*Walter and Eliza Hall Institute*

[†] *Givanna Putri and Felix Marsh-Wakefield contributed equally as co-last authors.*

## Abstract

GPU-accelerated Self-Organizing Map (SOM) implementations are among the most competitive options for large-scale SOM analysis, but growing dataset sizes increasingly challenge their practical use because workloads no longer fit cleanly within device-memory limits. We introduce FloatSOM, a SOM framework for scalable training and deployment that supports multi-GPU execution, out-of-memory disk-backed streaming, and novel topologies beyond regular lattices. We evaluate FloatSOM on 14 synthetic and real benchmark datasets together with controlled speed scaling benchmarks, and show that these improved topologies, combined with topology-aware hyperparameter fine-tuning, yield lower quantization error than current state-of-the-art SOM baselines. FloatSOM also sustains this performance at large scale with high-throughput distributed execution; in the largest benchmark, it trains a 1024-node SOM network on 1,000,000,000 samples with 50 features in 6.16 minutes on 8 GPUs across two separate high-performance-computing nodes.

## 1 Introduction

Self-Organizing Map (SOM), originally introduced by Kohonen (Kohonen, 1990), is an unsupervised machine-learning method that uses competitive learning to organize nodes such that they capture the topology of the data. In practice, this topology-preserving representation means SOMs are commonly used to map dataset topology, produce dimensionality-reduced visualizations, and conduct clustering at scale (Kangas et al., 1990). As dataset size and heterogeneity increase, the computational requirements of SOMs grow in both time and memory. Many current implementations, however, remain constrained to single-device workloads that must fit within video random-access memory (VRAM, GPU memory), with limited support for distributed compute, out-of-core execution, and modern GPU orchestration.

SOM topology presents a second limitation. Classical SOMs are traditionally trained on regular rectangular or hexagonal lattices because fixed grids make neighborhood definition, visualization, and optimization straightforward. However, these regular lattices also impose a strong geometric prior on the learned representation. Accordingly, prior work on dynamic, growing, and graph-structured SOM variants reflects a long-standing recognition that fixed lattices are not always the best match for irregular data geometry (Alahakoon et al., 2000; Vasighi and Amini, 2017; Kangas et al., 1990). However, these alternatives have generally not been developed or evaluated in the high-throughput, large-sample regime targeted by modern GPU-enabled applications and are broadly impractical for deployment at scale.

FloatSOM is designed to address these combined systems and topology limitations. Within FloatSOM, we implement distributed multi-GPU execution, out-of-memory disk-backed streaming, and scalable topology-flexible training beyond standard fixed lattices. Specifically, we implement highly scalable minimum-spanning-tree (MST) and relative-neighborhood-graph (RNG) topologies. We also evaluate multiple sampling strategies as a route to further computational acceleration and derive fine-tuned hyperparameter configurations across diverse datasets and recommended operating regimes. FloatSOM is suitable for both high-performance-computing (HPC) operation and consumer-grade desktop GPUs.

## 2 Related Work

Related work on practical SOM deployment spans software implementations, sampling-efficient training, topology design, and model selection.

### 2.1 Open-Source SOM Implementations and Systems

Open sourced SOM libraries range from lightweight to more performance-oriented implementations. MiniSom is a compact Python implementation of the classical online regime (Vettigli, 2018), whereas XPySOM is a Python-based batch SOM implementation designed for efficient GPU-backed execution (Mancini et al., 2020). At larger scales, Somoclu and GigaSOM provide mature parallel SOM systems for large workloads (Wittek et al., 2017; Kratochvíl et al., 2020). However, an important systems gap remains: XPySOM is limited to a single GPU and requires the full dataset to fit in VRAM, while GigaSOM emphasizes distributed CPU execution rather than distributed GPU training within a broadly used Python workflow. FloatSOM is intended to address that gap.

### 2.2 Sampling Methodologies for SOM Training

Classical online and batch SOM training schemes traditionally sample every data-point in every training iteration (Kohonen, 2001; Liu et al., 2023). Consequently, one obvious route to improving SOM speed and scalability is subsampling, which reduces the amount of data used to train the final SOM.

To date, numerous subsampling strategies have been proposed. Beyond naive random sampling, guided subsampling methods include hierarchical dynamic subset selection SOM (HDSSSOM), which concentrates computation on difficult or stale regions of the data (Wetmore et al., 2005), and adaptive sampling strategies for design-space exploration (Ito et al., 2016). However, systematically benchmarked and openly maintained implementations that compare full-data, random, and guided sampling within a modern high-performance SOM workflow remain limited.

## 2.3 SOM Lattice and Adaptive Topologies

Most practical SOM implementations retain regular rectangular or hexagonal lattices because they simplify neighborhood indexing, visualization, and vectorized updates (Kohonen, 1990, 2013). Hexagonal lattices are often preferred in the literature and serve as the regular-topology baseline in this manuscript (White and Kiester, 2008; Kohonen, 2013; Forest et al., 2020).

The SOM literature has also explored alternatives to fixed lattices, including dynamic maps and graph-structured neighborhoods (Vasighi and Amini, 2017; Spanakis and Weiss, 2016; Kangas et al., 1990; Jang et al., 2009). However, these alternatives have not been assessed on large-scale datasets and do not have implementations that are either publicly available or suitable for distributed GPU computation. None of the current distributed or GPU SOM implementations support these non-lattice topologies.

Relative Neighborhood Graphs (RNGs) (Toussaint, 1980) are of particular interest in presently; we return to the full rationale and implementation for RNG in Section 3.2.2.

## 2.4 Hyperparameter Optimization and Fair Comparison

SOM performance depends strongly on parameters such as map size, initialization, learning-rate schedule, and neighborhood schedule. Prior work shows that these choices can substantially affect observed performance (Akinduko et al., 2016; Forest et al., 2020). This makes comparisons based only on untuned defaults difficult to interpret.

More generally, modern machine-learning workflows increasingly rely on automated hyperparameter optimization rather than manual tuning alone. Tools such as Optuna provide bounded search over large hyperparameter spaces and can support multi-objective optimization, allowing parameter settings to be selected with respect to several benchmark criteria simultaneously rather than collapsed into a single score (Akiba et al., 2019).

# 3 Methods

FloatSOM combines four methodological components: sample selection, topology definition, SOM updates, and the compute architecture used to execute them. The framework supports random, full, and HDSSSOM sampling; rectangular, hexagonal, MST, and RNG topologies; and execution modes ranging from local GPU training to single-node and multi-node multi-GPU operation. Hyperparameter optimization is treated as a separate layer applied across these configurations. Figure 1 summarizes this design.

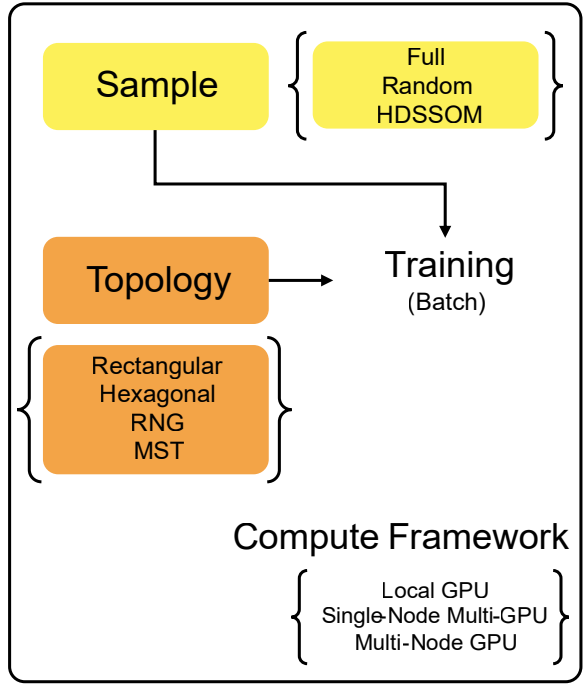

**Figure 1.** Schematic overview of the FloatSOM methods framing used in this manuscript. Sample selection and topology definition have configurable components that feed into the standard SOM training step, while the compute architecture determines how that same training procedure is executed in practice. The options shown here summarize the FloatSOM configurations discussed in the following subsections.

### 3.1 Sampling Selector Mathematics

This section formalizes the sampling policies evaluated in this manuscript.

Let the full dataset be $X = \{x_i\}_{i=1}^{N}$, where $N$ is the total number of samples in the dataset. Let $m$ denote the number of samples presented to the SOM in a given training iteration, or the sampling budget. This budget is either fixed directly or determined as a proportion $\rho$ of the dataset:

$$m = \begin{cases} m_0, & \text{if a fixed budget is specified,} \\ \max(1, \lfloor N\rho \rfloor), & \text{if a dataset proportion is used.} \end{cases} \qquad (1)$$

For index sets $\mathcal{I}_t^{(s)}$ at iteration $t$, the implemented full and random selectors are:

$$\mathcal{I}_t^{\text{full}} = \{1, \ldots, N\}, \qquad \mathcal{I}_t^{\text{random}} \sim \text{Unif}(\{I \subseteq \{1, \ldots, N\} : |I| = m\}), \ m < N. \qquad (2)$$

If $m \geq N$, random returns the full dataset (no subsampling). Otherwise, to avoid biasing training toward a single fixed subsample, the random selector redraws $\mathcal{I}_t^{\text{random}}$ from the entire dataset at every training iteration. The selected training subset is $X_t^{(s)} = \{x_i : i \in \mathcal{I}_t^{(s)}\}$. Because random subsampling draws uniformly from size-$m$ subsets, we are able to better obtain coverage over the full dataset.

HDSSSOM is implemented in multi-GPU-compatible form as an adaptive sampler that preferentially revisits difficult, under-trained, or stale samples (Wetmore et al., 2005).

## 3.2 Topology Definition

After initialization, neighborhood relations are defined by the selected topology. For regular-lattice baselines, we support both grid and hexagonal layouts. Consistent with prior SOM guidance, we treat hexagonal as the standard reference topology in this manuscript. We implement the hexagonal lattice topology in accordance with (Vettigli, 2018). For MST and RNG, neighborhood structure is derived from the current node-weight geometry using the methodologies below.

Note that regardless of the selected topology, FloatSOM uses the same node-weight initialization options. Initialization determines only the starting node weights; neighborhood relations are applied afterward according to the selected topology to establish node connections, keeping the initial state comparable across regular-lattice and graph-based runs.

### 3.2.1 MST Topology Implementation

The MST topology replaces fixed lattice neighborhood distance with graph hop distance on a minimum spanning tree built from current SOM nodes. For $P$ nodes in feature dimension $d$, the pairwise node-distance matrix is formed with the standard squared-distance Gram identity, which avoids 3D broadcast tensors and preserves $O(P^2 d)$ dense linear-algebra structure.

In a standard lattice SOM, neighborhood distance is determined by fixed node grid coordinates, which are assigned before training and are independent of the learned node-weight geometry in data space. MST instead uses the node weight vectors to locate the nodes in data space, then calculates the minimum spanning tree between those node locations (Kruskal, 1956). The node distances therefore determine which tree edges are selected, while the neighborhood distance used by the Gaussian update is the graph hop distance along the resulting tree. Topology-derived influence matrices are cached and refreshed according to the topology-refresh schedule.

At the beginning of training, the topology is refreshed every iteration: the previous tree edges are discarded, pairwise distances are recalculated from the updated node weights, and the MST is recomputed from those updated node locations. The refresh rate then decays as training proceeds, so topology updates become less frequent later in training. If the current iteration does not trigger recomputation, the previous graph state and cached influences are reused. Otherwise, the MST is recalculated and the influence cache is rebuilt.

---
**Algorithm 1** Dynamic MST topology update with refresh-triggered recomputation and cached influence reuse.

---
1: **Input:** node weights $W_t$, iteration $t$, topology-refresh policy
2: **Output:** topology state ($E_t$, $g_t$, cached influences)
3: Query the refresh policy for the current iteration
4: **if** no topology refresh is due **then**
5:     return previous topology state
6: **end if**
7: Compute pairwise squared node distances on GPU
8: Transfer distances to CPU and run Kruskal to obtain MST edges $E_t$
9: Build adjacency from $E_t$
10: Compute all-pairs graph hop distances $g_t$ with chunked GPU Floyd-Warshall
11: Deduplicate active radii and rebuild/update cached influence maps
12: Commit $E_t$, $g_t$, and cache state
13: return topology state

---

### 3.2.2 RNG Topology Implementation

Our RNG topology constructs a Relative Neighborhood Graph over current node distances using the standard RNG criterion (Toussaint, 1980). Unlike MST, RNG is not restricted to a single spanning-tree backbone, so multiple locally supported neighborhood relations can be retained. Conceptually, the absence of this restriction permits the RNG topology to represent both tree-like and denser mesh-like local structures simultaneously throughout the training process. Consequently, we hypothesise that this greater freedom in topology will allow RNG-based SOM maps to better represent a variety of distributions.

---

**Algorithm 2** Dynamic RNG topology update with refresh-triggered recomputation and cached influence reuse.

---
1: **Input:** node weights $W_t$, iteration $t$, topology-refresh policy
2: **Output:** topology state ($E_t, g_t$, cached influences)
3: Query the refresh policy for the current iteration
4: **if** no topology refresh is due **then**
5:     return previous topology state
6: **end if**
7: Compute pairwise squared node distances on GPU
8: Evaluate RNG candidate elimination in chunks using the blocker test
9: Retain surviving RNG edges $E_t$ and build adjacency
10: Compute all-pairs graph hop distances $g_t$ with chunked GPU Floyd-Warshall
11: Deduplicate active radii and rebuild/update cached influence maps
12: Commit $E_t$, $g_t$, and cache state
13: return topology state

---

## 3.3 Multi-GPU + Larger-Than-Memory Methodology and Implementation

FloatSOM is implemented for distributed GPU execution, with support for RAM (Random Access Memory, CPU memory) spilling and disk-backed execution when datasets exceed available VRAM and RAM, respectively.

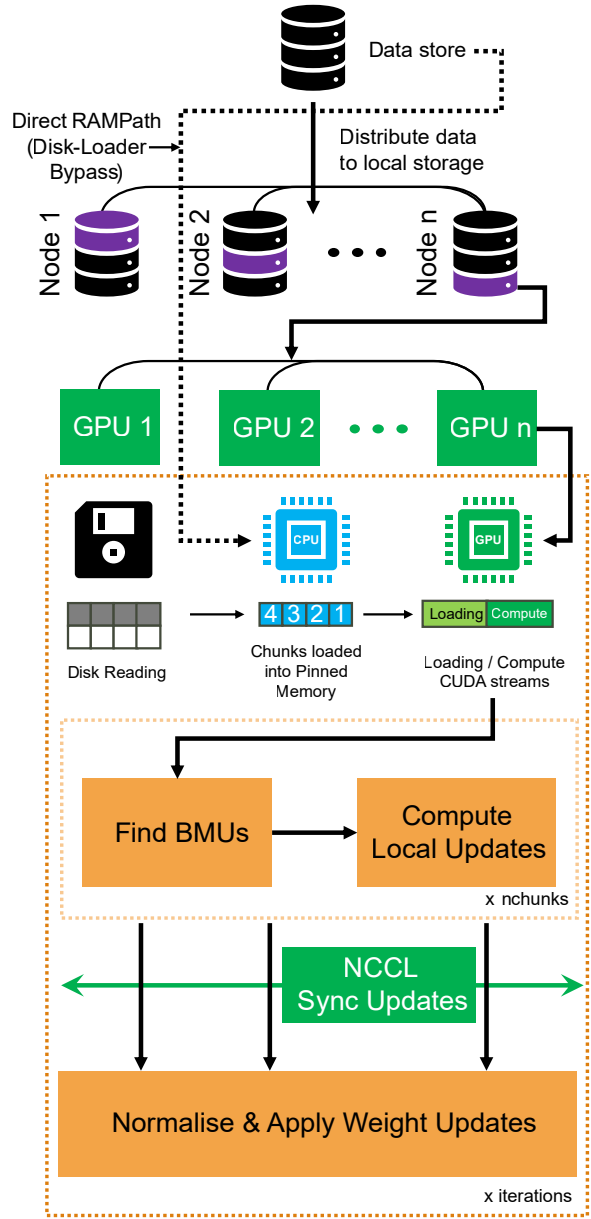

**Figure 2.** Multi-GPU data loading and NCCL synchronization schematic. In disk-backed mode, data are sharded to worker-local storage, loaded chunk-wise into pinned host memory, transferred to GPU with transfer overlapped with compute, processed locally, and synchronized by NCCL all-reduce before one weight update per iteration. In RAM mode, data are instead pre-sharded into worker-local CPU RAM and follow the same pinned-memory-to-GPU path without disk reads.

### 3.3.1 General Multi-GPU Logic

Distributed execution uses Ray actors with one GPU per worker and NCCL collectives for synchronous aggregation (Moritz et al., 2018). At each iteration, every worker processes its assigned shard locally in `n_chunks`, accumulates worker-local statistics across those chunks, and synchronizes once per iteration.

For worker $g \in \{1, \ldots, G\}$, let $X_{t,g}^{(s)}$ denote the worker-local shard of the selected iteration subset $X_t^{(s)}$. Let $j$ index SOM nodes, let $w_j^{(t)}$ denote the weight vector of node $j$ at iteration $t$, let $b(x)$ denote the best-matching unit (BMU) of sample $x$ under the current weights, let $h_{j,b(x)}^{(t)}$ denote the iteration-$t$ neighborhood influence between node $j$ and the BMU (Best Matching Unit) of $x$, and let $\eta_t$ denote the learning rate at iteration $t$. The local accumulators are:

$$U_j^{(g)} = \sum_{x \in X_{t,g}^{(s)}} \eta_t \, h_{j,b(x)}^{(t)} (x - w_j^{(t)}),$$
$$H_j^{(g)} = \sum_{x \in X_{t,g}^{(s)}} h_{j,b(x)}^{(t)}. \tag{3}$$

$U_j^{(g)}$ is worker $g$'s summed learning-rate-scaled displacement for node $j$, and $H_j^{(g)}$ is the corresponding summed neighborhood influence used to normalize that displacement.

Global synchronized accumulators are:

$$U_j = \sum_{g=1}^{G} U_j^{(g)},$$
$$H_j = \sum_{g=1}^{G} H_j^{(g)}. \tag{4}$$

In implementation, $U_j^{(g)}$ and $H_j^{(g)}$ are accumulated across the worker's `n_chunks` and synchronized once per iteration. After synchronization, $U_j$ and $H_j$ define the global update numerator and normalization denominator for node $j$. Under weighted normalization, the normalized update is given by $U_j/H_j$ up to numerical safeguards; optional momentum is then added to this normalized update before applying the resulting change to the node weight $w_j^{(t)}$.

### 3.3.2 Multi-GPU Implementation Details

As shown in Fig. 2, each worker uses a chunked loading path from CPU memory to GPU memory. In streaming mode, data are distributed to worker-local disk shards and then read chunk-by-chunk into pinned host memory before transfer to GPU; in RAM mode, data are pre-sharded into worker-local CPU RAM and follow the same pinned-memory path without disk reads. FloatSOM overlaps data transfer and compute across CUDA streams, uses JIT-compiled kernels (Just-In-Time) for the core BMU and update steps, and synchronizes worker-local accumulators once per iteration with NCCL all-reduce. Weights therefore remain resident on worker GPUs across iterations. For large topologies that would otherwise exceed VRAM, FloatSOM also supports spilling the associated graph-distance or influence structures from VRAM to system RAM.

### 3.3.3 Larger-Than-Memory Capable Topology Updates

Additional larger-than-memory support for topology updates is provided through topological chunking. Topological chunking applies the same idea to topology-side computations within each worker. When graph-distance or influence structures would otherwise exceed a worker's VRAM, those computations are tiled and evaluated in bounded pieces rather than materialized at once. This topology-side chunking is not depicted in the Fig. 2 data path, but it follows the same per-worker bounded-memory execution rule.

## 3.4 Multi-Objective Hyperparameter Optimization

We use Optuna as an automated multi-objective hyperparameter optimization tool to derive near-optimal performance and corresponding hyperparameters for each FloatSOM configuration under a given sampling and topology combination (Akiba et al., 2019).

# 4 Experimental Setup

We use two benchmark protocols: an Optuna quality benchmark and a speed scaling benchmark. The first evaluates algorithmic quality and tuned attainable performance, and the second evaluates runtime and distributed scaling behavior. All production Optuna and speed benchmarks reported in this manuscript were executed on Gadi at the National Computational Infrastructure (NCI), Australia, on gpuvolta nodes (HPC), using a consistent multi-GPU environment across runs.

## 4.1 Optuna benchmark protocol

We use Optuna-based multi-objective optimization to determine the best attainable performance and corresponding hyperparameters for each sampling (full and random) and topology (hexagonal, MST, and RNG) combination, optimizing jointly for train and holdout quantization error ($QE_T$ and $QE_H$). In this protocol, each dataset-topology-sampling configuration is optimized for 200 trials across 10 seeds under variant-appropriate search constraints. The reported runs used the standard in-memory batch path rather than the Ray-distributed execution stack. This corresponds to:

$$14_{\text{Datasets}} \times 10_{\text{Seeds}} \times 3_{\text{Topologies}} \times 2_{\text{SamplingMethods}} \times 200_{\text{Trials}} = 168{,}000 \text{ Runs}$$

Sampling comparisons in Section 5.2 compare full versus random paired analyses pooled across all topologies. Topology comparisons in Sections 5.3-5.4 use full sampling runs across hexagonal, MST, and RNG. Otherwise, the remaining analyses use full sampling. A separate focused HDSSSOM pilot is reported in Section 5.2 and summarized in Supplementary Table S2. This pilot additionally included the HDSSSOM sampling methodology alongside the full and random sampling methodologies.

### 4.1.1 Optuna benchmark datasets and preprocessing

The Optuna benchmark uses 14 synthetic and real scikit-learn datasets spanning low- and high-dimensional regimes (Pedregosa et al., 2011); dataset metadata are listed in Supplementary Table S1. Synthetic datasets are generated according to the random seed, whereas real datasets are loaded directly from sklearn. Across the Optuna protocol, inputs are standardized feature-wise to zero mean and unit variance using `StandardScaler`, then deterministically permuted and split into fixed 70/30 train-holdout partitions. We retain both partitions to evaluate both training-set representation and transfer to unseen samples.

### 4.1.2 Optuna quality metrics

Our primary quality metric is Quantization Error ($QE$) (Kohonen, 2001). We report $QE$ on both the training partition ($QE_T$) and the holdout partition ($QE_H$). The training value reflects how well the fitted SOM represents the samples it was trained on, whereas the holdout value evaluates the same trained map after projecting previously unseen samples onto it. The holdout partition is therefore not used during fitting; it is used only to evaluate the trained map, analogous to a test set in conventional model-training workflows.

Balanced $QE$, denoted $QE_B$, is defined as the mean of $QE_T$ and $QE_H$. $QE_B$ is therefore a composite endpoint that weights data representation fidelity (train) and generalisability (holdout) equally. Note that in the Optuna runs, $QE_T$ and $QE_H$ are optimized jointly as a two-objective vector, with $QE_B$ only calculated *post hoc*.

### 4.2 Speed scaling benchmark protocol

The speed scaling benchmark evaluates runtime and distributed scaling behavior in different compute and algorithm configurations. Speed scaling is evaluated with runs across $G \in \{1, 2, 4, 8\}$ GPUs under a series of fixed scaling protocols. Runtime summaries are computed from repeated executions per configuration and reported as both absolute training time and efficiency relative to the corresponding 1-GPU comparison. We compute scaling efficiency as $E_G = (T_1/T_G)/G \times 100\%$, where $T_1$ is the 1-GPU runtime and $T_G$ is the runtime on $G$ GPUs. A value of 100% corresponds to ideal linear scaling; lower values indicate that communication, orchestration, or I/O overhead has reduced the realized speedup, while values above 100% can occur when additional GPUs also change the memory or data-staging regime. These scaling runs use the Ray-orchestrated distributed execution layer built on top of the standard FloatSOM training path (Moritz et al., 2018). Accordingly, the scaling figures in Sections 6.1-6.2 and the runtime/scaling comparison reported later against XPySOM should be interpreted as distributed-execution results rather than the in-memory Optuna path.

For scaling-efficiency calculations, when a single-GPU baseline was missing at a given axis value due to timeout, we estimated that baseline by local linear extrapolation from the last available 1-GPU point on the same curve. That is, runtime was assumed to scale proportionally with the axis variable for the extrapolation step (for example, doubling sample count or doubling dimensionality doubles the estimated 1-GPU runtime).

Topology speed comparisons include hexagonal, MST, and RNG, with harmonized workload settings so ratios isolate topology-associated runtime effects. A dedicated random versus full comparison, run on $G \in \{1, 2, 4\}$ GPUs, is additionally included to isolate sampling-specific runtime effects independently of the multi-GPU scaling runs.

#### 4.2.1 Scaling benchmark datasets

The speed benchmark uses synthetic random matrices with uniform values in $[0, 1]$. No train-holdout split is used here because the objective is runtime rather than generalization, so all generated data are used for training. We use a shared default configuration of $10^7$ samples, 50 dimensions, a $32 \times 32$ SOM grid, and 10 training iterations, and then vary one workload axis at a time: sample count ($10^6$–$10^9$), feature dimension (50–5000), or grid side length (8–64). The exact axis values used for each scaling panel are provided in Supplementary Table S3.

Additional CPU and RAM resources attached to each GPU are scaled linearly with GPU count while keeping the software environment and benchmark procedure consistent across runs. For scaling benchmarks, each run is timed out if it exceeds 30 minutes of wall time. Per-GPU resource allocation is fixed at one NVIDIA V100 (32 GB VRAM), 12 CPU cores, 90 GB system RAM, with one full node comprising 4 GPUs and their associated CPUs, and with 400 GB of associated local disk storage.

## 4.3 Hyperparameter Tuning and Stability

To quantify parameter-tuning benefit, we performed an explicit paired analysis between the tuned configuration and the XPySOM untuned default reference. For each sampling mode and topology combination, we first extracted the parameter settings from the best-performing Optuna runs under the benchmark objective for that combination. We then distilled these per-seed best-performing tuned settings into deployable default configurations by taking the mean of numeric parameters and the mode of categorical parameters.

The stability analysis focuses on four tuned hyperparameters that govern SOM training dynamics. These settings determine how the map is initialized and how updates evolve over training: the initial radius sets the early neighborhood scale around each BMU, the initialization method sets the starting node weights, the radius decay type controls the shift from broad global organization toward local refinement, and the momentum-use parameter determines whether each update retains part of the previous update direction.

### 4.3.1 Tuned Configuration versus Untuned Reference Analysis

This fixed derived configuration was then rerun and paired against the untuned default XPySOM (hexagonal) reference within matched dataset, seed, and dataset-split units.

Dataset-wise tuned configuration versus untuned reference summaries use the same forest-plot convention. The global overall summary pools all matched tuned configuration versus untuned reference pairs across datasets and applies the same paired $t$-test and confidence-interval construction to that pooled paired set.

### 4.3.2 Hyperparameter stability and dataset-type stratification

Hyperparameter stability is important because it determines whether a method can be deployed reliably without bespoke retuning. We therefore extracted the top-ranked tuned Optuna trial separately for each seed and compared the recovered hyperparameter values across seeds. We define hyperparameter stability as the variation in these recovered settings. For numeric parameters, stability is measured by the relative difference between runs, $|a-b|/\max(|a|,|b|,\varepsilon)$, where $a$ and $b$ are the values of a given parameter for the two compared seeds and $\varepsilon = 10^{-12}$. These relative differences are then averaged across numeric parameters, with lower values indicating higher stability. For categorical parameters, stability is defined as the mismatch rate across the same seed pairs. We report both mean per-parameter stability and an equal-weight overall stability summary across datasets.

## 4.4 Statistical analysis

All Optuna comparisons use matched pairs within dataset, seed, and split units to control for substantial between-run heterogeneity. Within each matched unit, trials were ranked by the target metric, the top five were retained, and each condition was summarized by the median of those retained trials, yielding a top-$k$ summary with $k=5$ intended to estimate near-optimal attainable performance under a fixed Optuna budget. Paired effects were then computed as simple condition differences, with negative values favoring the first condition for lower-is-better metrics. Dataset-level and global summaries are shown as forest plots with 95% confidence intervals from two-sided paired one-sample $t$-tests; top-$k$ sensitivity analyses are provided in the Supplementary figures.

# 5 Results

## 5.1 XPySOM calibration (Equivalence)

Under matched-configuration XPySOM versus FloatSOM calibration on hexagonal $QE$ (Fig. S3), the two implementations perform equivalently, with no detected $QE$ differences. Accordingly, we treat hexagonal FloatSOM batch as a valid proxy for XPySOM in the benchmarks that follow. The runtime differences are attributable to FloatSOM's JIT kernels, which incur a small startup cost. This overhead is progressively amortized as workload size increases, after which FloatSOM runs faster than XPySOM on larger datasets.

## 5.2 Comparison of Different Sampling Methods

We first report a focused HDSSSOM pilot as an elimination comparison rather than as part of the broader sampling benchmark. This pilot used a smaller, more limited configuration than the later full versus random analysis and is summarized in Supplementary Table S2.

Under that pilot configuration, HDSSSOM was materially worse than full sampling. In the hexagonal view shown in Fig. 3, full outperformed HDSSSOM in all 50 paired comparisons (10 datasets × 5 seeds; 50 wins, 0 losses, 0 ties), with dataset-level median balanced-$QE$ improvements ranging from 2.5% to 209.7% and a global median improvement of 38.7%. Given these large differences, we focused on the full and random sampling methodologies for the remainder of the manuscript.

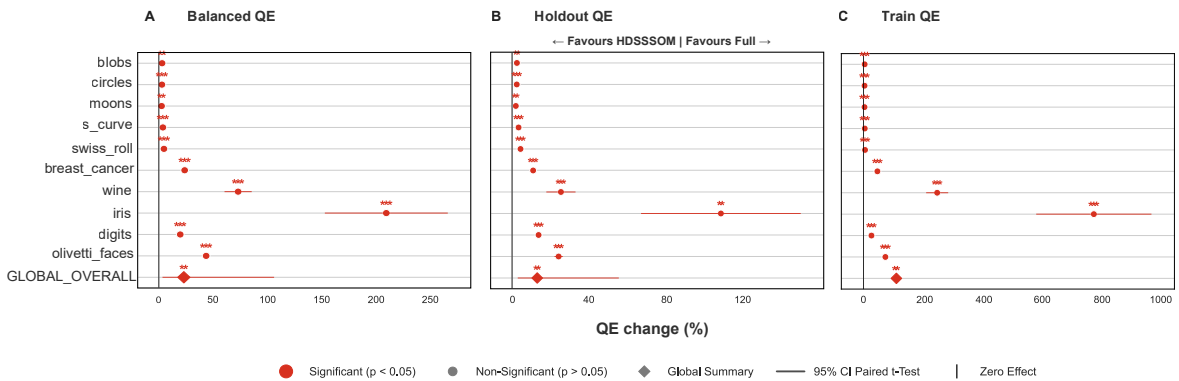

**Figure 3.** HDSSSOM screening pilot on $QE_B$ (hexagonal topology): full vs HDSSSOM, using the smaller pilot configuration summarized in Supplementary Table S2 (10 datasets, 5 seeds, with all other pilot settings held fixed). Panels report dataset-matched paired top-$k$ within-unit medians plus dataset-level paired-effect summaries (Section 4.3). Forest whiskers denote 95% paired $t$-test confidence intervals around the mean paired effect.

The full versus random analysis in Fig. 4 was generated from matched hexagonal Optuna runs in which the sampling selector was switched from full to random. Effects were computed within matched seed-specific units and then summarized across seeds. Fig. 4A-C summarize the matched full versus random paired effects for balanced, holdout, and train $QE$, respectively, showing that full sampling generally provides equal or better $QE$ than random sampling. Notably, this full versus random separation is much smaller than the full versus HDSSSOM pilot effect shown in Fig. 3; in most dataset-level comparisons, the full versus HDSSSOM improvement is at least twice as large - as evidenced by the difference in axis scales between Fig. 3 and Fig. 4.

We find the effectiveness of random sampling relative to full sampling is scale-dependent. Above 10,000 samples, paired $QE$ differences are not meaningfully detected, whereas in smaller datasets the random arm shows higher variability and less stable outcomes (Fig. 4D-F). This is consistent with reduced per-iteration sample support under random subsampling (Fig. 4D-F). In this benchmark, the $> 10{,}000$ regime is therefore a useful practical proxy for more stable random sampling behavior.

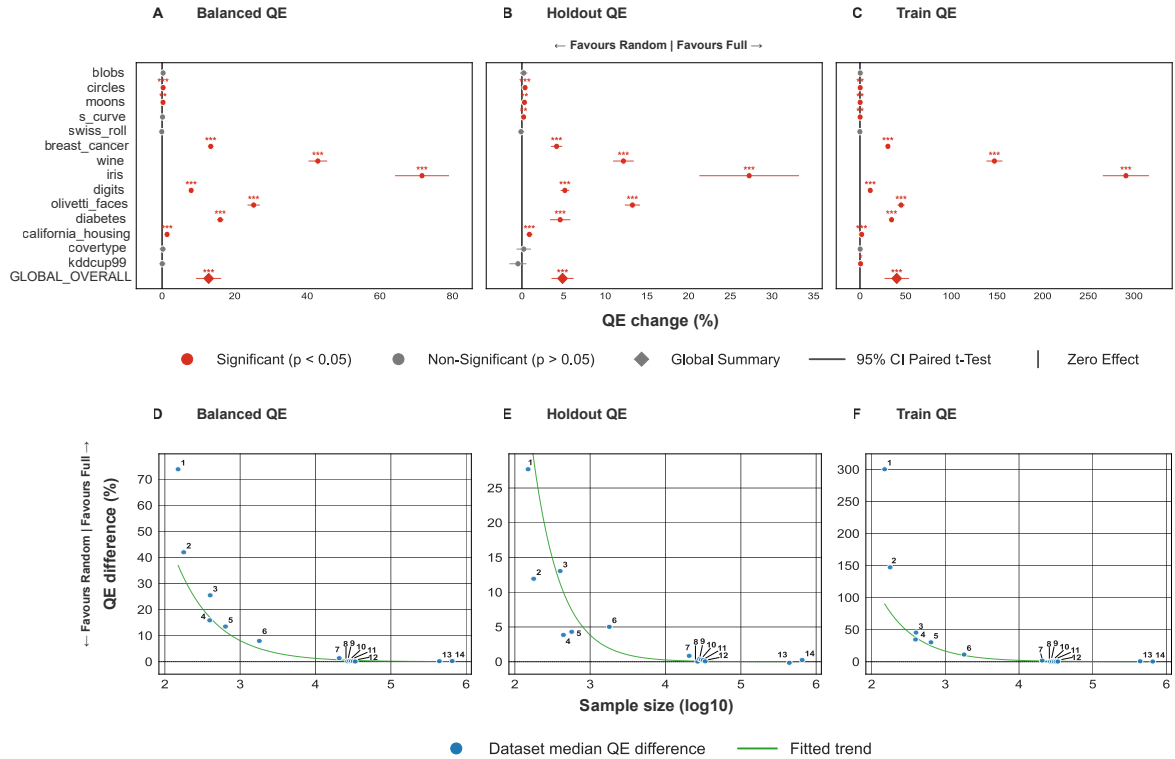

**Figure 4: Full vs Random Outcomes and Dataset-Size Regression (Hexagonal)**

**Figure 4.** Full versus random sampling under the hexagonal topology. Panels A-C report paired full-versus-random effects for $QE_B$, $QE_H$, and $QE_T$. Panels D-F show the corresponding dataset-size regressions for the same three metrics, with numbered datasets keyed in Supplementary Table S1. In the larger-dataset regime observed here (>10,000 samples), little paired $QE$ separation is detected; at smaller dataset scales, random is more variable and less stable.

Nevertheless, full sampling remains the best sampling strategy for optimal $QE$ results. Accordingly, all remaining analyses reported below rely on full sampling unless specified.

## 5.3 Topology Results

Topology comparisons are reported with $QE$-only metrics, with the Optuna hexagonal batch setting as the primary regular-topology baseline; Fig. 5 provides a qualitative illustration of the neighborhood structures produced by hexagonal, MST, and RNG.

**Figure 5: Representative Topology on sklearn Circles**

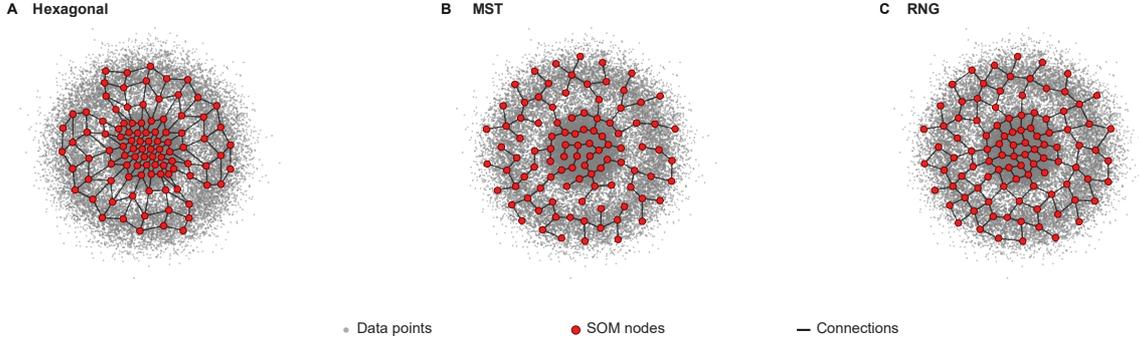

**Figure 5.** Representative neighborhood node and connection overlays for default XPySOM hexagonal, MST, and RNG runs on a 30,000 data-point synthetic sklearn circles dataset.

### 5.3.1 MST

We report dataset-wise paired improvement summaries (hexagonal over MST) with the same reporting logic as Section 5.1 in Fig. 6.

Overall, MST outperforms matched hexagonal on balanced QE (Fig. 6A), indicating a net advantage across train and holdout performance. This aggregate gain is driven more clearly by train QE (Fig. 6C), while holdout QE is more mixed across datasets (Fig. 6B) and shows no clear overall holdout advantage. The overall paired t-test p-values are balanced QE (p=1.12e-05), holdout QE (p=0.15), and train QE (p=0.0064).

**Figure 6: Hexagonal vs MST Across QE Metrics**

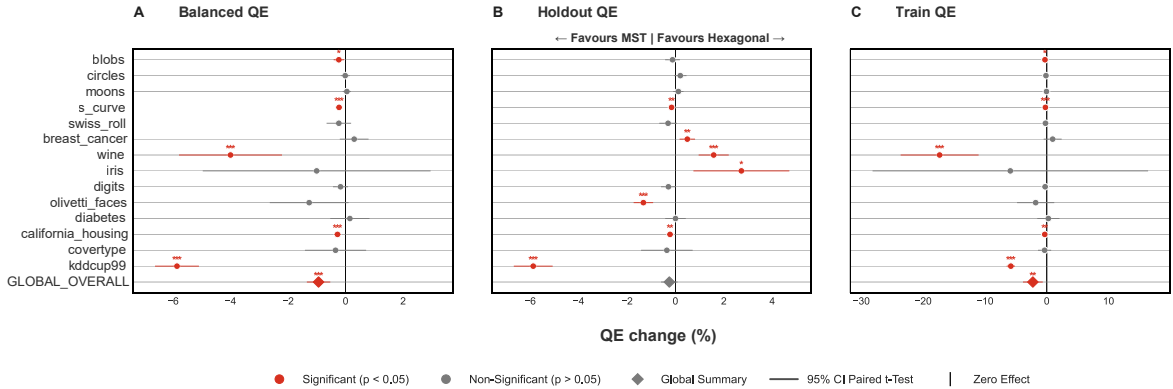

**Figure 6.** Hexagonal versus MST topology on $QE$ metrics under full sampling only. Panels A-C report paired full sampling only $QE$ effects for $QE_B$, $QE_H$, and $QE_T$ across the available full sampling datasets. Forest whiskers denote 95% paired $t$-test confidence intervals around the mean paired effect.

### 5.3.2 RNG

To evaluate RNG topology performance, we reuse the paired reporting logic on hexagonal versus RNG, again centered on $QE_B$ with $QE_H$ and $QE_T$ in Fig. 7.

RNG has lower QE than matched hexagonal on the reported QE metrics (Fig. 7A-C), with overall paired t-test p-values of balanced QE (p=7.4e-10), holdout QE (p=0.0232), and train QE (p=4.69e-06).

The main trend in Fig. 7 is that RNG improves on hexagonal most clearly in balanced QE and especially in train QE, with the separation most apparent in the real and larger datasets where the added flexibility of the graph neighborhood appears more useful than the fixed regular lattice.

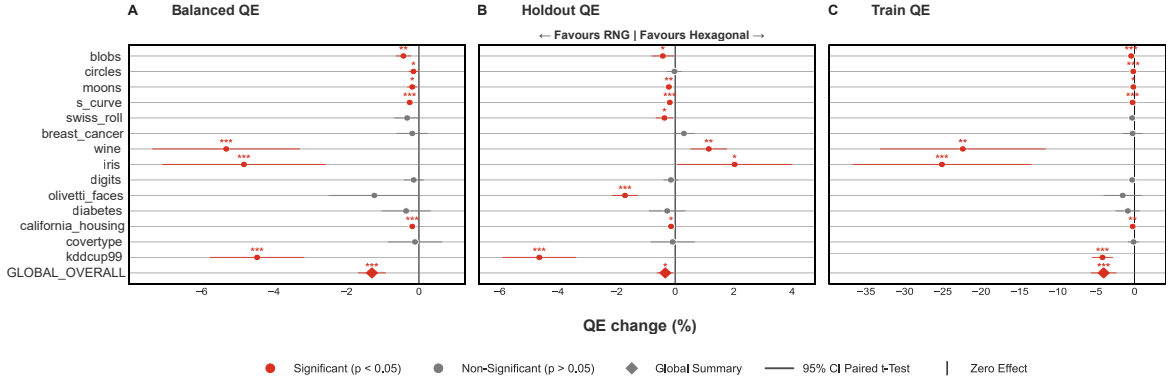

**Figure 7.** Hexagonal versus RNG topology on $QE$ metrics under full sampling only. Panels A-C report paired full sampling only $QE$ effects for $QE_B$, $QE_H$, and $QE_T$ across the available full sampling datasets. Forest whiskers denote 95% paired $t$-test confidence intervals around the mean paired effect.

### 5.4 Hyperparameter Tuning and Stability

#### 5.4.1 Performance Gains from Hyperparameter Tuning

To quantify the practical benefit of deploying tuned settings, we compare tuned and untuned configurations in Fig. 8 using pooled topology-level $QE$ summaries. Topology-stratified versions of the same comparison are provided in Figs. S8-S10. The tuned configurations are the Section 4.3 derived settings, and the untuned reference is the default XPySOM configuration. Broadly, tuning yields substantial and significant $QE$ improvement.

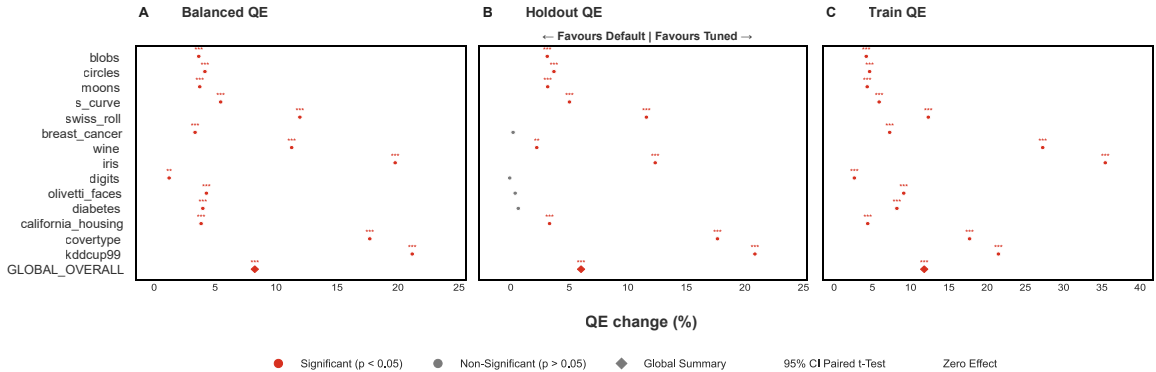

**Figure 8.** Tuned configuration versus untuned reference $QE$ comparison across $QE_B$, $QE_H$, and $QE_T$, pooled across all topology runs under the matched pairing keys. Positive values indicate the tuned configuration outperforms the untuned reference; the global overall row pools all matched tuned configuration/untuned reference pairs across datasets. Forest whiskers denote 95% paired *t*-test confidence intervals around the mean paired effect.

At the pooled overall level, the paired summaries across all matched tuned configuration/untuned reference pairs also favor tuning for all three metrics, consistent with the per-dataset pattern in Fig. 8. The same tuning pattern is observed across topologies, indicating that tuning affects all topology families rather than a single-architecture artifact.

### 5.4.2 Hyperparameter Stability Across Topology and Sampling

To assess whether the derived hyperparameters are robust across runs and datasets, we compare the within-topology seed-to-seed tuned-parameter drift (Section 4.3.2). Overall, MST and RNG exhibits less variability in performance compared to hexagons (Fig. 9A using full dataset, 9B using random sampling). Across these panels, full sampling is generally the more stable setting. Mirroring the sample-size-dependent $QE$ performance, Fig. 9C shows that random sampling hyperparameter stability also improves with increasing sample size.

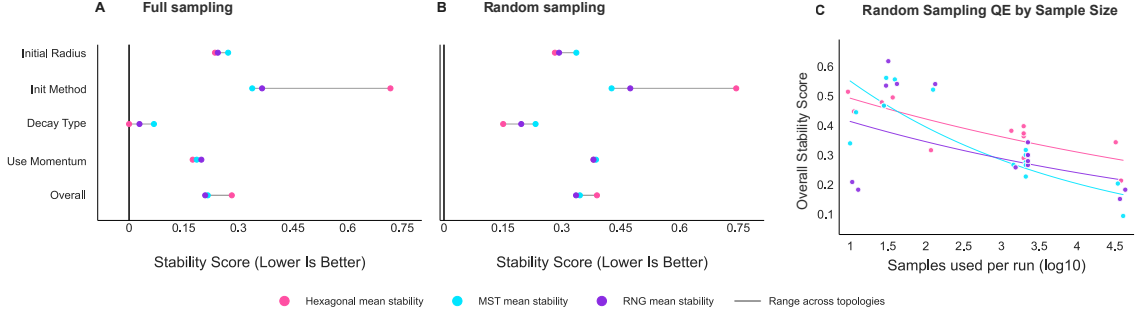

**Figure 9.** Hyperparameter stability by sampling mode. A: selected parameter stability under full sampling for hexagonal, MST, and RNG topologies (lower stability score is better). B: selected parameter stability under random sampling for the same topologies. C: dataset-size stability regression under random sampling, using the selected parameter stability score against sample size (log10) across the included topology families.

# 6 Speed Scaling

We next examine three aspects of FloatSOM runtime performance: the cost of full relative to random sampling, the scaling behavior obtained with additional GPUs, and the runtime implications of MST and RNG relative to a hexagonal topology.

## 6.1 Random versus Full Sampling Runtime

We report sample scaling runtime comparisons for full versus random sampling using the synthetic datasets as outlined in Section 4.2.1. We stratify by hexagonal, MST, and RNG, for 1, 2, and 4 GPUs (Fig. 10).

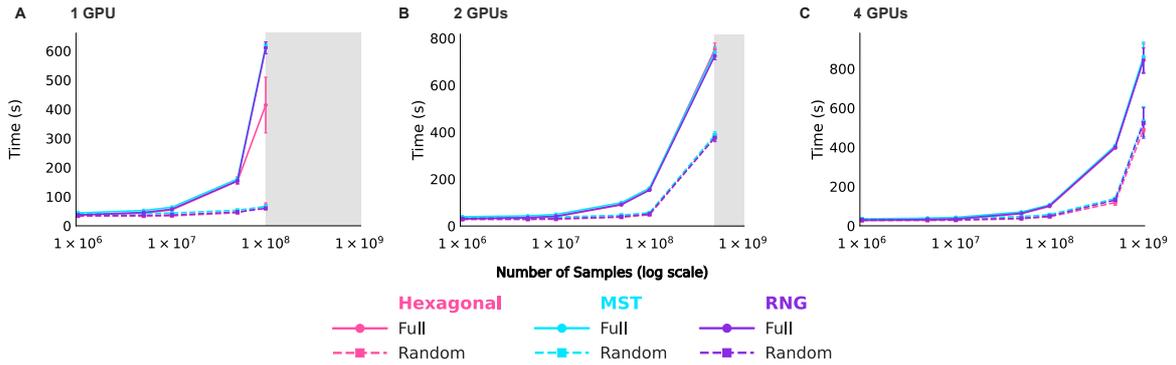

**Figure 10.** Sample scaling runtime comparison of full versus random sampling across $G \in \{1, 2, 4\}$ GPUs (A,B,C). Curves report mean wall-clock training time (s) under harmonized settings; error bars denote $\pm 1$ standard deviation across $n = 3$ repeated runs per configuration. Color encodes topology (hexagonal, MST, RNG), and line style encodes sampling mode (full vs. random). Shaded x-axis regions indicate sample-size ranges that could not be run in that panel relative to the shared axis maximum due to timeouts. Lower values indicate faster execution.

Across topologies, random sampling is faster than full sampling across all 1-, 2-, and 4-GPU comparisons, with similar proportional reductions at a given dataset size. With more GPUs, larger datasets can also be processed before timing out, although the runtime advantage of random over full narrows at the largest sample counts.

For the 1-GPU random sampling runs, the last successful 1-GPU run was at 100M samples, the same number of samples as run on full samples, despite random being much faster. We reason that the failed subsequent runs on larger samples were due to the transition from RAM operation to disk-backed operation. Even in random sampling mode, the implementation still stages the full dataset initially and loads full chunks before the subsampler selects the training samples. Therefore, the relevant cost is no longer only the reduced number of selected samples. We therefore treat the 1-GPU random failure at 500M samples as a disk-mode systems limitation rather than as evidence against the general random versus full runtime ordering.

## 6.2 Multi-GPU topology scaling and Larger-Than-Memory context

To examine parallel scaling, we benchmarked FloatSOM under full sampling across $G \in \{1, 2, 4, 8\}$ GPUs using workload-scaling benchmarks that extend from standard in-memory settings to larger workloads that exceed available memory. Across these increasingly demanding loads, runtime and efficiency show a broadly consistent scaling pattern, with stable behaviour regardless of topology (Figs. 11 and S11).

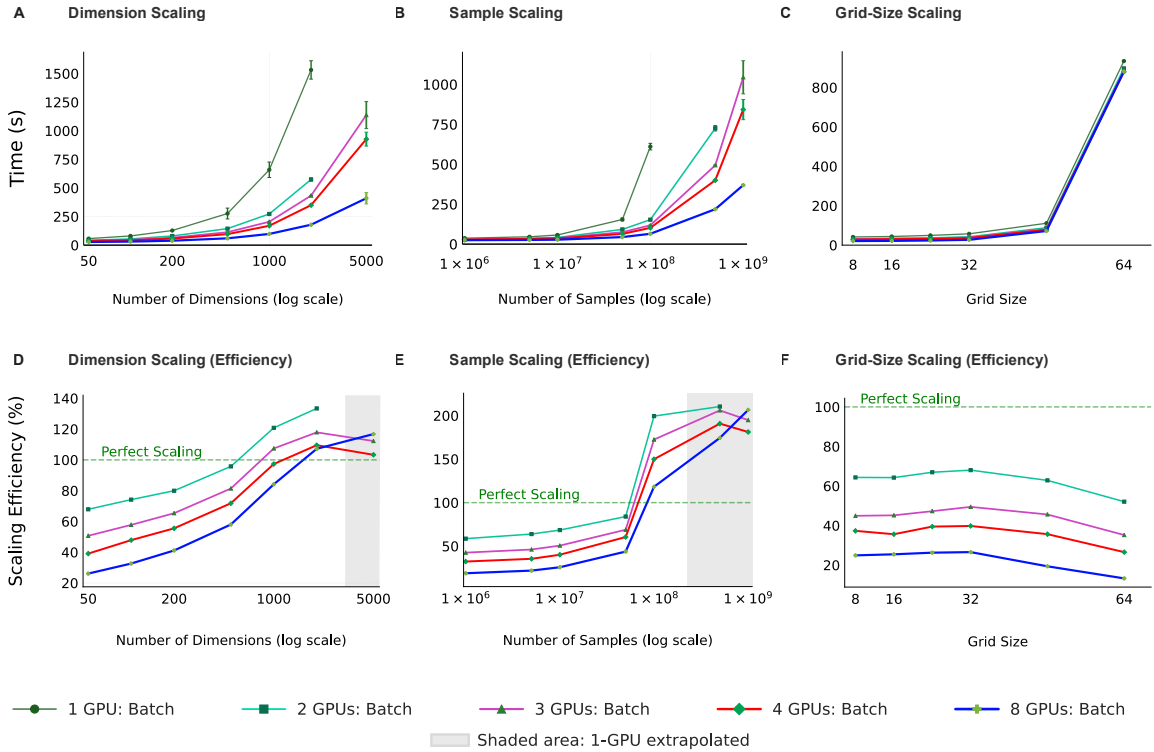

**Figure 11:** Multi-GPU Full-Batch Scaling (RNG)

**Figure 11.** Multi-GPU scaling across $G \in \{1, 2, 4, 8\}$ GPUs. Panels A-C show runtime (s) for dimension, sample, and grid size scaling workloads, respectively. Panels D-F show scaling efficiency for the same workloads, computed from the single-GPU baseline and the corresponding $G$-GPU runtime. Runtime error bars denote $\pm 1$ standard deviation across $n = 3$ repeated runs per configuration; the 100% efficiency reference line indicates ideal linear scaling.

### 6.2.1 GPU Scaling and Larger-Than-Memory Runtime Performance

Fig. 11 shows that increasing GPU count improves performance in the sample scaling regime by increasing parallel throughput and delaying the transition to disk-backed execution. In the sample scaling benchmark, the 500,000,000 sample dataset requires disk backing under the 2-GPU configuration, whereas the 8-GPU configuration remains in RAM mode until the 1,000,000,000 sample dataset; when staging is still required, the disk-to-GPU path is distributed across more nodes. The 8-GPU RNG configuration processes 1,000,000,000 samples in 369.41 s (6.16 min). Note that this speed includes the time required to transfer data from shared storage to node-local shards, alongside the overhead associated with operating across multiple HPC nodes.

The grid size scaling panel proves to be the main exception: at the largest tested grid size (64), runtime shortens from 934.01 s (15.57 min) on 1 GPU to only 880.83 s (14.68 min) on 8 GPUs, a 5.69% reduction, indicating that once map-size/topology-refresh costs dominate, additional GPUs contribute little extra speedup.

### 6.2.2 GPU Scaling Efficiency

We next consider GPU efficiency under strong scaling, relative to ideal linear scaling. At smaller dataset sizes, efficiency is lower. As workload size increases, efficiency rises sharply and in some regions exceeds 100%. When a direct 1-GPU baseline was unavailable at a given axis value, the efficiency denominator was constructed by local linear extrapolation from the last available 1-GPU point on that curve (Section 4.2), so some values should be interpreted with care if the underlying 1-GPU runtime is nonlinear over that range. Fig. 11 shows the scaling for the RNG topology, while Fig. S11 shows the corresponding supplementary hexagonal and MST outputs, which follow the same overall pattern.

## 6.3 Topology Runtime Comparisons

With that systems context in place, we next compare topology runtimes across hexagonal, MST, and RNG configurations on 8 GPUs (Fig. 12).

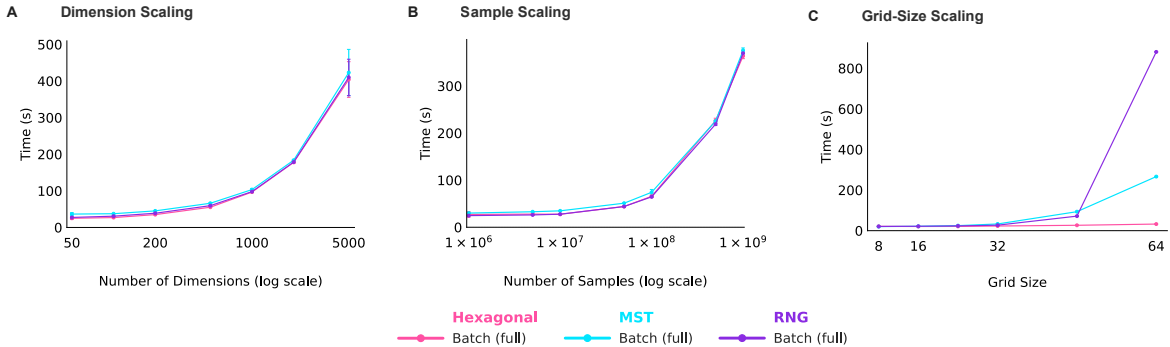

**Figure 12.** Topology runtime comparison at fixed $G = 8$ GPUs under full sampling. Panels A-C report mean wall-clock runtime (s) for dimension, sample, and grid size scaling workloads, respectively, with topology traces for hexagonal, MST, and RNG. Error bars denote $\pm 1$ standard deviation across $n = 3$ repeated runs per configuration. The largest-axis 8-GPU topology runtime summaries are listed in Supplementary Table S11.

In Fig. 12A-B, the topologies scale similarly as input complexity and data volume increase: even at the largest tested axis values, the maximum pairwise runtime spread remains modest at dimension scaling (4.70% at 5,000 dimensions) and sample scaling (3.26% at 1,000,000,000 samples).

However, when the grid itself is enlarged in Fig. 12C, topology-dependent runtime differences become readily evident. At the largest tested grid size (grid size 64), the 8-GPU mean runtimes are 32.54 s (0.54 min) for hexagonal, 266.45 s (4.44 min) for MST, and 880.83 s (14.68 min) for RNG, corresponding to 8-GPU MST and RNG runtimes that are 8.19x and 27.07x the hexagonal runtime, respectively.

# 7 Final FloatSOM RNG Comparison with XPySOM

Fig. 13 demonstrates the $QE$ gains from topology choice and tuning persist in deployment against default hexagonal XPySOM. (Mancini et al., 2020).

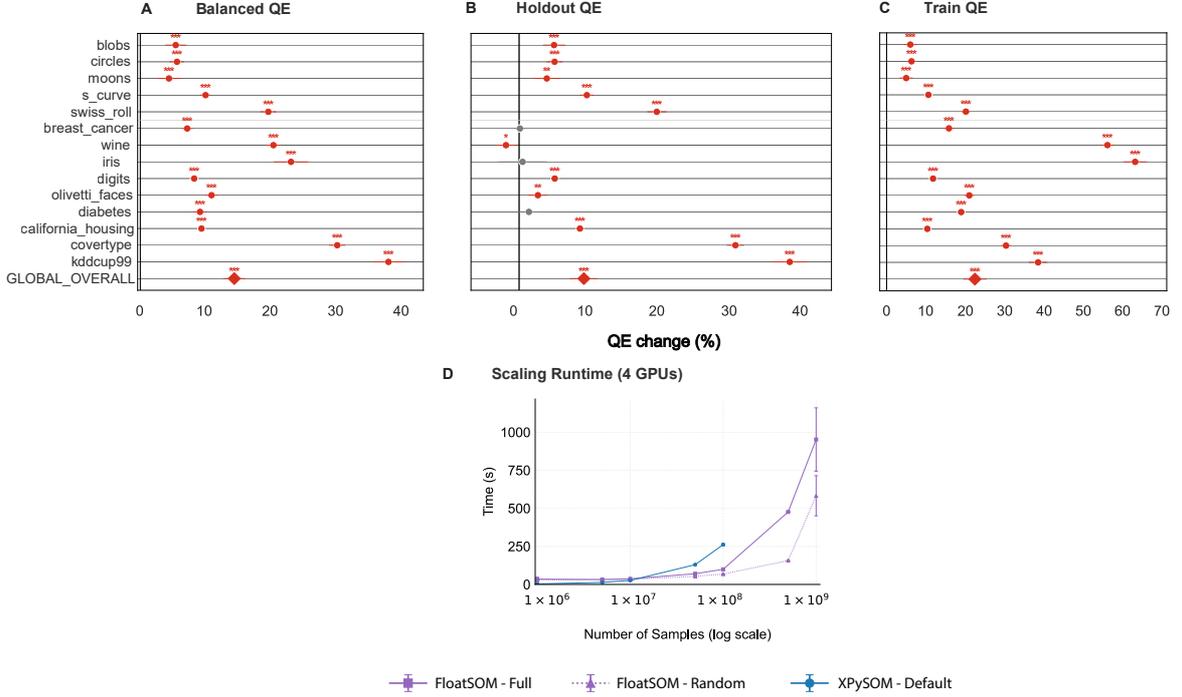

**Figure 13.** Integrated deployment comparison of default hexagonal XPySOM versus tuned FloatSOM RNG. Panels A-C compare $QE_B$, $QE_H$, and $QE_T$ using the untuned hexagonal XPySOM baseline against matched tuned FloatSOM RNG full sampling runs. Panel D provides the scaling/runtime context for the same comparison.

At the overall level, Fig. 13 shows median percentage improvements of $QE_B$ (14.5%); $QE_H$ (9.1%); and $QE_T$ (22.5%) for tuned FloatSOM RNG relative to default hexagonal XPySOM, capturing the combined deployment effect of topology choice and tuning on $QE$.

For the default hexagonal XPySOM reference in Fig. 13, workloads beyond the $10^8$-sample case were not processed because they exceeded available VRAM and XPySOM requires the full dataset to be loaded into memory. Taken together, Fig. 13 shows the point at which the workload size is large enough to warrant the extra startup overhead incurred by tuned FloatSOM RNG: tuned FloatSOM RNG delivers better $QE$ than the default hexagonal XPySOM baseline, while also running faster and scaling to larger workloads.

## 8 Discussion

This manuscript introduces FloatSOM as a GPU-oriented SOM framework that combines topology flexibility, out-of-memory execution, and distributed multi-GPU training in a single implementation. The study also provides a unified empirical analysis of sampling strategy, graph-based versus fixed-lattice topology, topology-aware hyperparameter tuning, and workload scaling across synthetic and real benchmarks. Taken together, these results clarify how algorithmic and systems choices jointly shape SOM quality and runtime under practical deployment constraints.

## 8.1 Sampling tradeoff (random versus full)

The sampling trade-off is strongly scale dependent. In smaller datasets, random subsampling produces less stable outcomes, whereas above 10,000 samples paired $QE$ differences between full and random are not meaningfully detected. Full sampling is therefore the safer choice when stability is the priority, while random sampling is better viewed as a throughput-oriented option at larger scales. In the larger-than-memory regime, this advantage narrows because random still requires each worker-local chunk to be read before subsampling, so it reduces compute more directly than dataset I/O.

## 8.2 Topology Comparisons (MST and RNG)

Globally, both MST and RNG outperform the fixed hexagonal topology in these comparisons, with RNG showing the strongest overall $QE$ results. One interpretation is that this ordering reflects increasing structural flexibility across the topology families, with the more flexible graph-based neighborhoods conforming more effectively to the underlying data distribution than the fixed lattice baseline (Kohonen, 2013; Kangas et al., 1990; Toussaint, 1980). The denser connected structure available under RNG may also provide additional regularization, because nodes can receive information from more neighbors during updating rather than being limited to a single tree path. This could support more precise local updates, although the present benchmark does not isolate that mechanism directly.

## 8.3 Tuning benefit under matched defaults

Across the matched Fig. 8 comparisons, tuned configurations consistently outperform untuned reference settings. Hyperparameter tuning should therefore be treated as part of the method configuration rather than as optional post-processing.

The key implication is that topology choice and hyperparameter choice are coupled. Gains remain positive across hexagonal, MST, and RNG, so tuning is not confined to a single topology. A practical deployment strategy therefore requires topology-aware tuning.

## 8.4 Hyperparameter stability and dataset-type interpretation

The stability results suggest that graph topologies are recovered more consistently than the fixed-lattice baseline. Hexagonal maps cannot rebuild connectivity once initialized, whereas MST and RNG recompute connectivity from the evolving node-weight geometry. It may also help explain why the graph topologies are better suited to higher-dimensional, non-synthetic datasets (Kohonen, 2013; Kangas et al., 1990).

## 8.5 Systems implications and limits

Distributed execution should therefore be interpreted as workload-dependent rather than as a uniform multiplicative speedup. At small problem sizes, Ray startup, communication, and staging overheads appear to dominate, consistent with the lower efficiencies observed at the low end of the scaling curves. At larger workloads, the sharp rise in efficiency likely reflects both increased parallel throughput and a change in memory regime, where some multi-GPU configurations remain in RAM while the 1-GPU baseline has already entered disk-backed execution. Apparent efficiencies above 100% should therefore be interpreted as reflecting this systems transition rather than literal super-linear compute scaling.

Overall, these results support a practical deployment strategy that uses RNG with topology-aware tuned defaults. For optimal speed, consider using the largest number of GPUs available, especially if this enables data to be kept in RAM. Sampling should be chosen by scale: full for smaller datasets when stability is critical, and random as a throughput-oriented option in the larger-dataset regime where optimal $QE$ is not essential. For workloads dominated by very large grid size scaling, MST remains a reasonable alternative to RNG and Hexagonal.

This manuscript presents FloatSOM as a unified large-scale SOM framework that combines a novel graph-based topology with sampling options, optimised hyperparameters, and distributed out-of-memory GPU execution. This integrated design supports practical SOM deployment at scale, where topology choice, sampling, quantization performance, and systems constraints can be managed together rather than in isolation.

# 9 Acknowledgements

This work was supported by computational resources provided by the Australian Government through the National Computational Infrastructure (NCI) under the ANU Merit Allocation Scheme.

We also acknowledge the computational services provided by the University of Bern, the University of Sydney, and the Walter and Eliza Hall Institute.

We thank Prof. Hanna Suominen for her input and advice.

# 11  Supplementary Tables (End Matter)

Supplementary Table S1. Dataset metadata and numbered point key for the Figure 4 sampling mode analysis.

| dataset_index | dataset | dimension_count | sample_size | dataset_type |
| --- | --- | --- | --- | --- |
| 1 | iris | 4 | 150 | real |
| 2 | wine | 13 | 178 | real |
| 3 | olivetti_faces | 4096 | 400 | real |
| 4 | diabetes | 10 | 442 | real |
| 5 | breast_cancer | 30 | 569 | real |
| 6 | digits | 64 | 1797 | real |
| 7 | california_housing | 8 | 20640 | real |
| 8 | blobs | 2 | 30000 | synthetic |
| 9 | circles | 2 | 30000 | synthetic |
| 10 | moons | 2 | 30000 | synthetic |
| 11 | s_curve | 3 | 30000 | synthetic |
| 12 | swiss_roll | 3 | 30000 | synthetic |
| 13 | kddcup99 | 41 | 494021 | real |
| 14 | covertype | 54 | 581012 | real |

Supplementary Table S2. HDSSSOM pilot configuration summary for Figure 3.

| field | value |
|---|---|
| pilot purpose | Initial HDSSSOM screening before the broader sampling comparison |
| configuration scope | Smaller hexagonal Optuna pilot |
| datasets | `swiss_roll`, `moons`, `circles`, `blobs`, `s_curve`, `breast_cancer`, `wine`, `iris`, `digits`, `olivetti_faces` |
| dataset count | 10 |
| seed count | 5 |
| sampling methods present | full, random, HDSSSOM |
| topology | `hexagonal` |
| algorithm family in campaign | `colors` |
| optimization split setup | `evaluation-split=both`, yielding `QE_H` and `QE_T`; Figure 3 reports paired `QE_B` |
| trials per scenario | 200 |
| main text figure slice | hexagonal / full vs HDSSSOM |

Supplementary Table S3. Exact axis values used for the sample count, feature dimension, and grid size scaling benchmarks in Section 4.2.1.

| scaling axis | exact values |
|---|---|
| sample count | $10^6$, $5 \times 10^6$, $10^7$, $5 \times 10^7$, $10^8$, $5 \times 10^8$, $10^9$ |
| feature dimension | 50, 100, 200, 500, 1000, 2000, 5000 |
| grid side length | 8, 16, 24, 32, 48, 64 |

**Supplementary Table S4. FloatSOM versus XPySOM calibration $QE$ summary for the MST topology path.** The `dataset_index` column matches the numbered points in Supplementary Figure S1 panel D. See `assets/tables/supp_xpysom_calibration_qe_mst.tsv`.

**Supplementary Table S5. FloatSOM versus XPySOM calibration $QE$ summary for the RNG topology path.** The `dataset_index` column matches the numbered points in Supplementary Figure S2 panel D. See `assets/tables/supp_xpysom_calibration_qe_rng.tsv`.

**Supplementary Table S6. FloatSOM versus XPySOM calibration $QE$ summary for the hexagonal topology path.** The `dataset_index` column matches the numbered points in Supplementary Figure S3 panel D. See `assets/tables/supp_xpysom_calibration_QE_Hexagonal.tsv`.

**Supplementary Table S7. Paired topology comparison p-values for hexagonal versus MST and hexagonal versus RNG across balanced QE, holdout QE, and train QE.** Rows list metric/dataset entries, including the OVERALL row. The MST and RNG columns report p-values using the manuscript reporting convention. See `assets/tables/supp_table_topology_hex_vs_mst_rng_pvalues.tsv`.

**Supplementary Table S8. Figure 13 deployment comparison percent summary for tuned FloatSOM RNG versus default hexagonal XPySOM across $QE_B$, $QE_H$, and $QE_T$.** Rows list per-dataset and `GLOBAL_OVERALL` entries with the plotted median percent change and 95% confidence interval. See `assets/tables/supp_table_figure_13_xpysom_rng_deployment_summary.tsv`.

**Supplementary Table S9. Supplementary Figure S12 deployment comparison per-

cent summary for tuned FloatSOM hexagonal versus default hexagonal XPySOM across $QE_B$, $QE_H$, and $QE_T$. Rows list per-dataset and `GLOBAL_OVERALL` entries with the plotted median percent change and 95% confidence interval. See `assets/tables/supp_table_s13_xpysom_hexagonal_deployment_summary.tsv`.

**Supplementary Table S10. Supplementary Figure S13 deployment comparison percent summary for tuned FloatSOM MST versus default hexagonal XPySOM across** $QE_B$**,** $QE_H$**, and** $QE_T$**.** Rows list per-dataset and `GLOBAL_OVERALL` entries with the plotted median percent change and 95% confidence interval. See `assets/tables/supp_table_s14_xpysom_mst_deployment_summary.tsv`.

**Supplementary Table S11. Figure 12 topology runtime summary at the largest common 8-GPU axis value for the dimension, sample, and grid size scaling workloads.** Rows report the plotted 8-GPU mean runtimes for hexagonal, MST, and RNG, together with the fastest and slowest topology at that axis value and the maximum pairwise runtime spread. See `assets/tables/supp_table_figure_12_topology_runtime_summary.tsv`.

## 12 Supplementary Figures (End Matter)

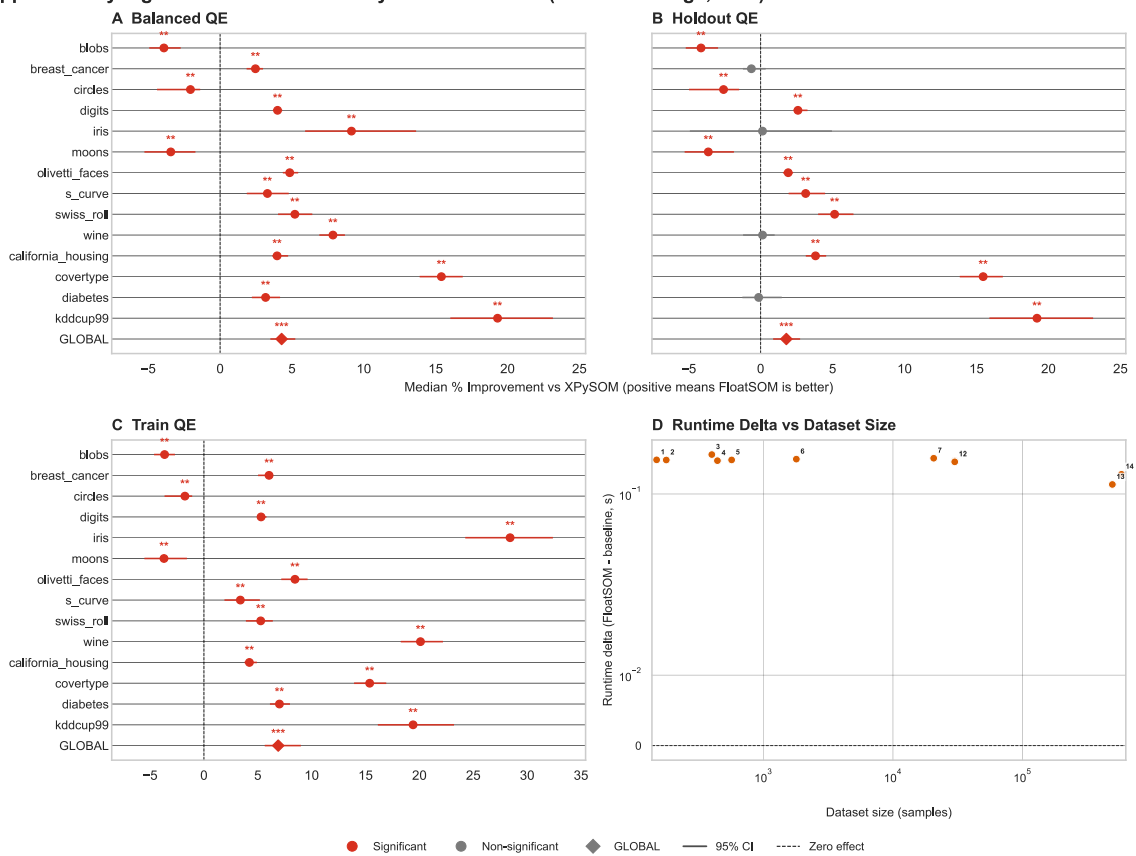

Supplementary Figure S1. FloatSOM versus XPySOM calibration under default settings for the MST topology path. Panels A-C report paired $QE$ effects for $QE_B$, $QE_H$, and $QE_T$. Panel D reports dataset-level median runtime deltas against dataset size, where each numbered dot is the median matched-seed value of `FloatSOM time - XPySOM time`; negative values favor FloatSOM and positive values favor XPySOM. The point numbers map to Supplementary Table S4. Forest whiskers denote 95% paired $t$-test confidence intervals around the mean paired effect.

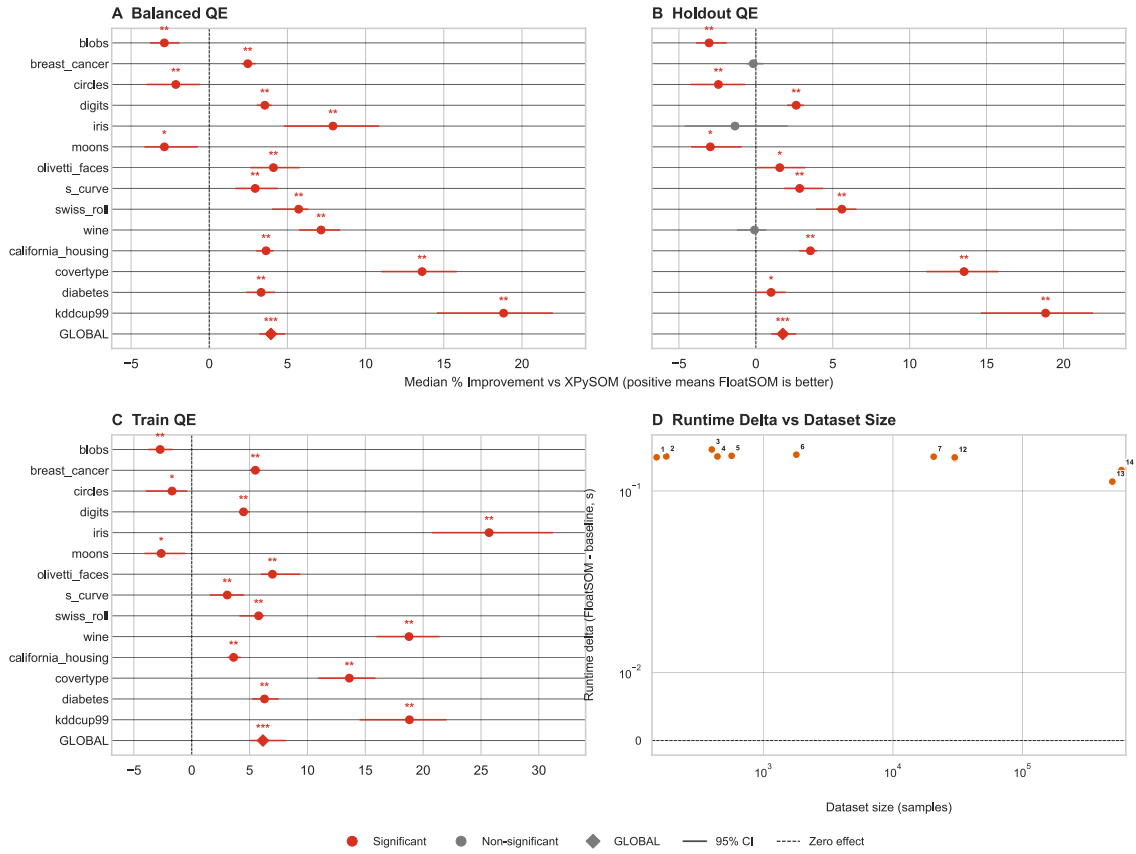

Supplementary Figure S2. FloatSOM versus XPySOM calibration under default settings for the RNG topology path. Panels A-C report paired $QE$ effects for $QE_B$, $QE_H$, and $QE_T$. Panel D reports dataset-level median runtime deltas against dataset size, where each numbered dot is the median matched-seed value of `FloatSOM time - XPySOM time`; negative values favor FloatSOM and positive values favor XPySOM. The point numbers map to Supplementary Table S5. Forest whiskers denote 95% paired $t$-test confidence intervals around the mean paired effect.

**Supplementary Figure S3: FloatSOM vs XPySOM Calibration (Default Settings, Hexagonal)**

Supplementary Figure S3. FloatSOM versus XPySOM calibration under default settings for the hexagonal topology path. Panels A-C report paired $QE$ effects for $QE_B$, $QE_H$, and $QE_T$. Panel D reports dataset-level median runtime deltas against dataset size, where each numbered dot is the median matched-seed value of `FloatSOM time - XPySOM time`; negative values favor FloatSOM and positive values favor XPySOM. The point numbers map to Supplementary Table S6. Forest whiskers denote 95% paired $t$-test confidence intervals around the mean paired effect.

**Supplementary Figure S4: MST vs RNG Across QE Metrics**

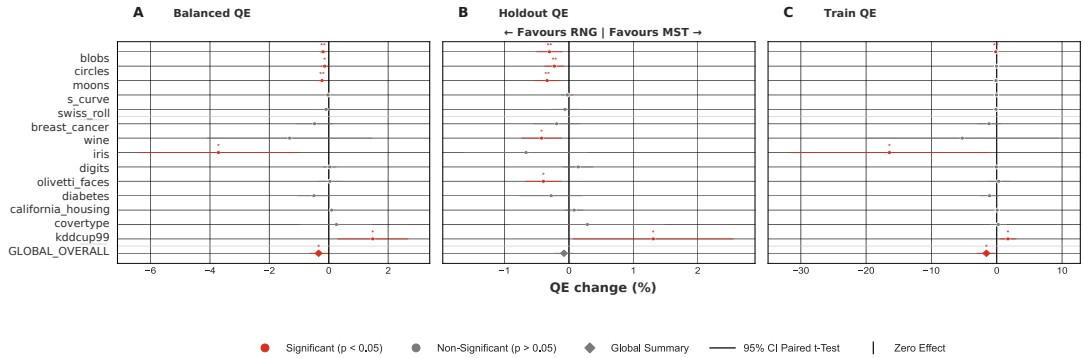

Supplementary Figure S4. MST versus RNG topology on $QE$ metrics under full sampling only. Panels A-C report paired full sampling only $QE$ effects for $QE_B$, $QE_H$, and $QE_T$ across the available full sampling datasets. Forest whiskers denote 95% paired $t$-test confidence intervals around the mean paired effect.

**Supplementary Figure S5: Hexagonal vs MST Sensitivity Across QE Metrics**

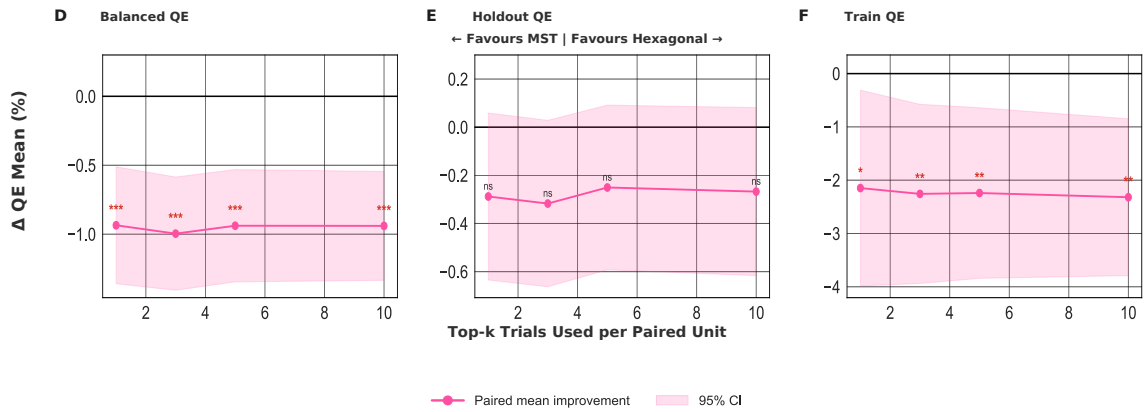

Supplementary Figure S5. Hexagonal versus MST topology sensitivity under full sampling only. Panels A-C report the matched top-$k$ paired sensitivity analysis for $QE_B$, $QE_H$, and $QE_T$.

**Supplementary Figure S6: Hexagonal vs RNG Sensitivity Across QE Metrics**

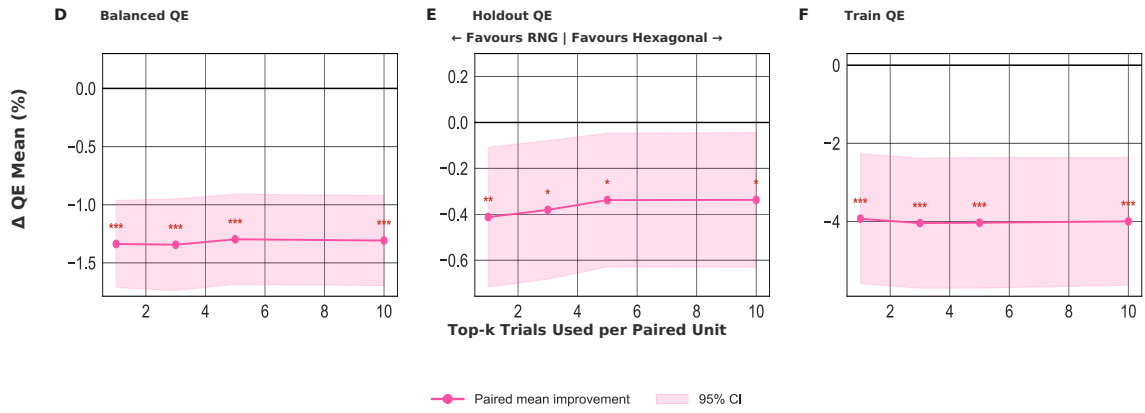

Supplementary Figure S6. Hexagonal versus RNG topology sensitivity under full sampling only. Panels A-C report the matched top-$k$ paired sensitivity analysis for $QE_B$, $QE_H$, and $QE_T$.

**Supplementary Figure S7: MST vs RNG Sensitivity Across QE Metrics**

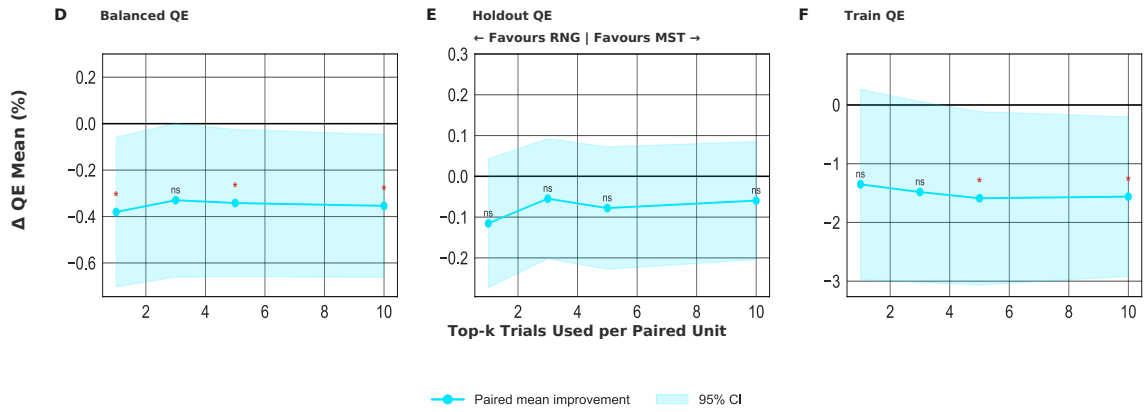

Supplementary Figure S7. MST versus RNG topology sensitivity under full sampling only. Panels A-C report the matched top-$k$ paired sensitivity analysis for $QE_B$, $QE_H$, and $QE_T$.

**Supplementary Figure S8: Tuned vs True Default Across QE Metrics (Hexagonal)**

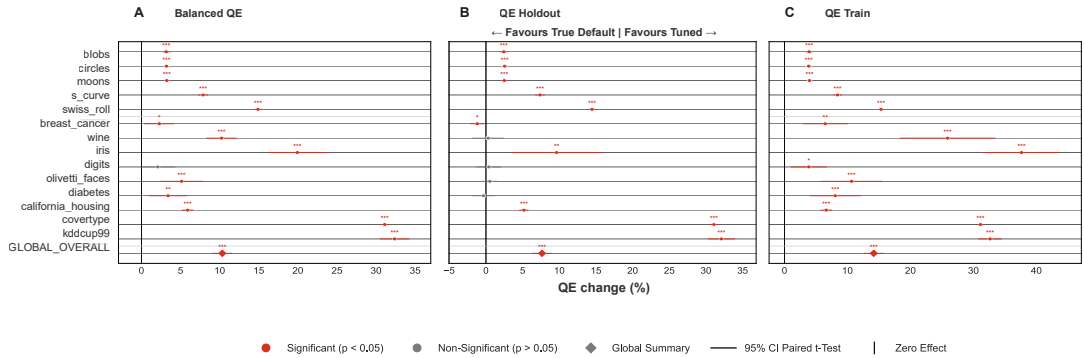

Supplementary Figure S8. Tuned configuration versus untuned reference $QE$ comparison for the hexagonal topology only, across $QE_B$, $QE_H$, and $QE_T$ under the matched pairing keys. Positive values indicate the tuned configuration outperforms the untuned reference. Forest whiskers denote 95% paired $t$-test confidence intervals around the mean paired effect.

**Supplementary Figure S9: Tuned vs True Default Across QE Metrics (MST)**

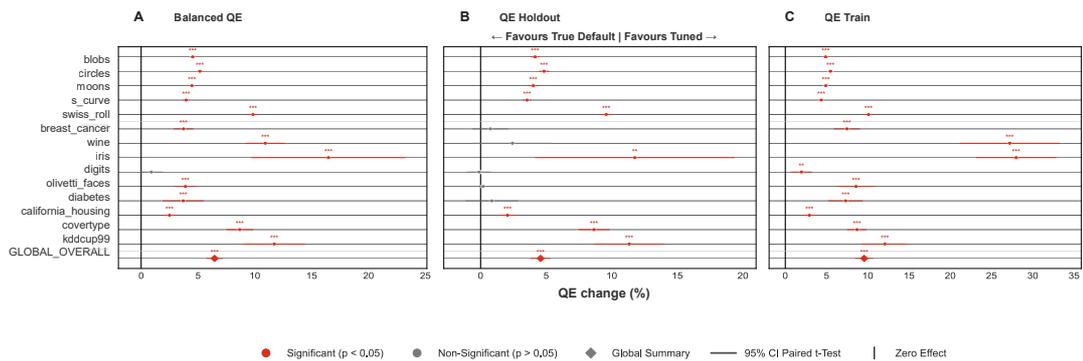

Supplementary Figure S9. Tuned configuration versus untuned reference $QE$ comparison for the MST topology only, across $QE_B$, $QE_H$, and $QE_T$ under the matched pairing keys. The tuned configuration is derived from the Optuna-selected settings by taking the mean of numeric parameters and the mode of categorical parameters across seeds. Positive values indicate the tuned configuration outperforms the untuned reference. Forest whiskers denote 95% paired $t$-test confidence intervals around the mean paired effect.

**Supplementary Figure S10: Tuned vs True Default Across QE Metrics (RNG)**

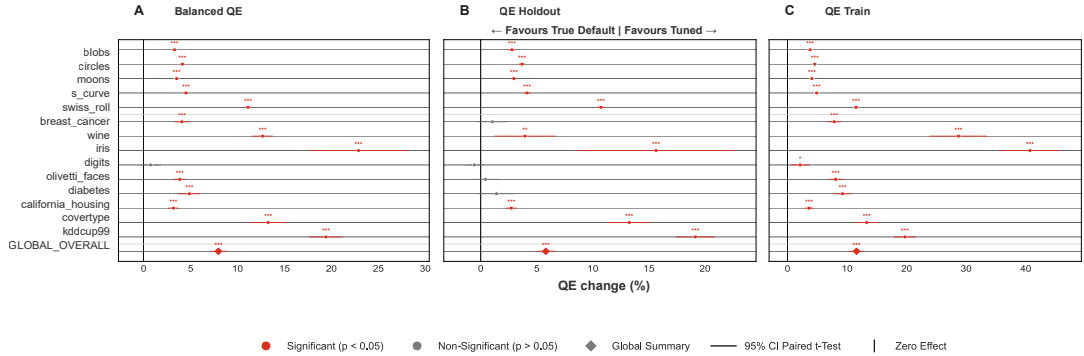

Supplementary Figure S10. Tuned configuration versus untuned reference $QE$ comparison for the RNG topology only, across $QE_B$, $QE_H$, and $QE_T$ under the matched pairing keys. The tuned configuration is derived from the Optuna-selected settings by taking the mean of numeric parameters and the mode of categorical parameters across seeds. Positive values indicate the tuned configuration outperforms the untuned reference. Forest whiskers denote 95% paired $t$-test confidence intervals around the mean paired effect.

**Supplementary Figure S11: Per-Topology GPU Scaling Performance (MST and Hexagonal)**

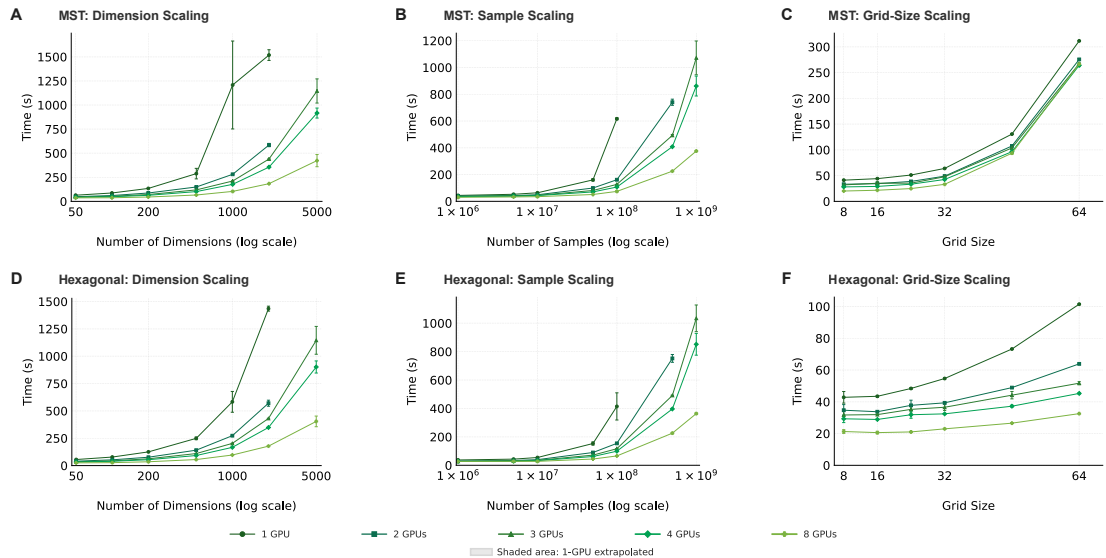

Supplementary Figure S11. Full GPU-count scaling context for MST and hexagonal under matched full sampling settings. Panels A-C show MST runtime across dimension, sample, and grid size scaling workloads; panels D-F show the corresponding hexagonal runs. Within each panel, curves correspond to $G \in \{1, 2, 4, 8\}$ GPUs and report mean wall-clock runtime (s) with $\pm 1$ standard-deviation error bars across $n = 3$ repeated runs per configuration.

**Supplementary Figure S12: XPySOM (Default) v FloatSOM Hexagonal**

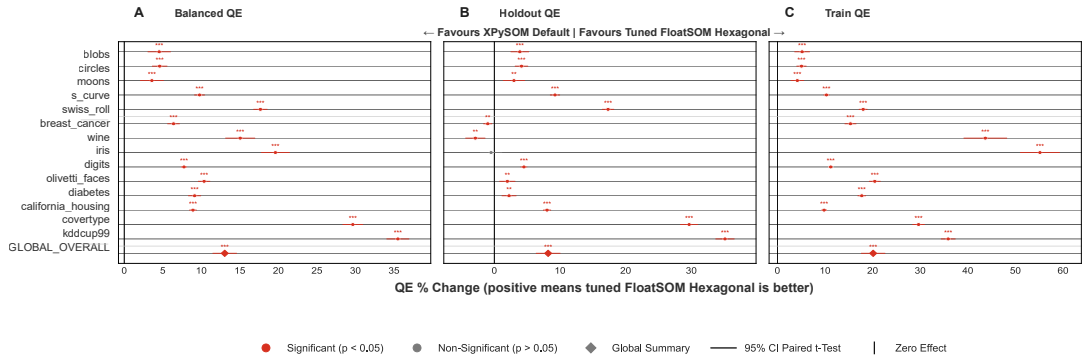

Supplementary Figure S12. Deployment comparison of default hexagonal XPySOM versus tuned FloatSOM hexagonal. Panels A-C report paired $QE$ effects for $QE_B$, $QE_H$, and $QE_T$ under the matched dataset/seed comparison keys. Positive values indicate tuned FloatSOM hexagonal outperforms default hexagonal XPySOM. Forest whiskers denote 95% paired $t$-test confidence intervals around the mean paired effect. Per-dataset and `GLOBAL_OVERALL` panel summaries are listed in Supplementary Table S9.

**Supplementary Figure S13: XPySOM (Default) v FloatSOM MST**

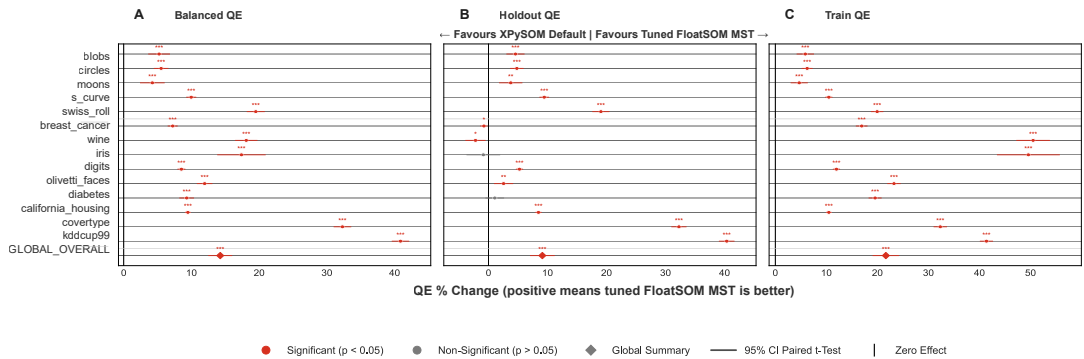

Supplementary Figure S13. Deployment comparison of default hexagonal XPySOM versus tuned FloatSOM MST. Panels A-C report paired $QE$ effects for $QE_B$, $QE_H$, and $QE_T$ under the matched dataset/seed comparison keys. Positive values indicate tuned FloatSOM MST outperforms default hexagonal XPySOM. Forest whiskers denote 95% paired $t$-test confidence intervals around the mean paired effect. Per-dataset and `GLOBAL_OVERALL` panel summaries are listed in Supplementary Table S10.